\documentclass[twocolumn,aps,prc,longbibliography,superscriptaddress,nofootinbib,showpacs,floatfix]{revtex4-1}

\usepackage{graphicx}	
\usepackage{xspace}	

\newcommand{\pt}    {\mbox{$p_T$}\xspace}
\newcommand{\Rgamma}{\mbox{$R_{\gamma}$}\xspace}

\newcommand{\Npart} {\mbox{$N_{\rm part}$}\xspace}
\newcommand{\Nqpart}{\mbox{$N_{\rm qp}$}\xspace}
\newcommand{\Ncoll} {\mbox{$N_{\rm coll}$}\xspace}
\newcommand{\Nincl} {\mbox{$N^ {{\rm incl}     }_\gamma$}\xspace}
\newcommand{\Npi}   {\mbox{$N^ {\pi^0,{\rm tag}}_\gamma$}\xspace}

\newcommand{\sqsn}  {\mbox{$\sqrt{s_{_{NN}}}$}\xspace}
\newcommand{\ee}    {\mbox{$e^+e^-$}\xspace}
\newcommand{\ef}    {\mbox{$\langle\varepsilon_\gamma f\rangle$}\xspace}
\newcommand{\AuAu}  {\mbox{Au$+$Au}\xspace}
\newcommand{\PbPb}  {\mbox{Pb$+$Pb}\xspace}
\newcommand{\pp}    {\mbox{$p$$+$$p$}\xspace}
\newcommand{\ptee}  {\mbox{$p^{ee}_T$}\xspace}

\newcommand{\ptmin} {\mbox{$p_T^{\rm min}$}\xspace}
 
\usepackage{nicefrac}

\begin{document}

\title{Centrality dependence of low-momentum direct-photon production in 
Au$+$Au collisions at $\sqrt{s_{_{NN}}}=200$~GeV}

\newcommand{\abilene}{Abilene Christian University, Abilene, Texas 79699, USA}
\newcommand{\augie}{Department of Physics, Augustana College, Sioux Falls, South Dakota 57197, USA}
\newcommand{\banaras}{Department of Physics, Banaras Hindu University, Varanasi 221005, India}
\newcommand{\barc}{Bhabha Atomic Research Centre, Bombay 400 085, India}
\newcommand{\baruch}{Baruch College, City University of New York, New York, New York, 10010 and The Graduate Center, City University of New York, New York, New York 10016 USA}
\newcommand{\bnlcoll}{Collider-Accelerator Department, Brookhaven National Laboratory, Upton, New York 11973-5000, USA}
\newcommand{\bnlphys}{Physics Department, Brookhaven National Laboratory, Upton, New York 11973-5000, USA}
\newcommand{\caucr}{University of California - Riverside, Riverside, California 92521, USA}
\newcommand{\charlesczech}{Charles University, Ovocn\'{y} trh 5, Praha 1, 116 36, Prague, Czech Republic}
\newcommand{\chonbuk}{Chonbuk National University, Jeonju, 561-756, Korea}
\newcommand{\ciae}{Science and Technology on Nuclear Data Laboratory, China Institute of Atomic Energy, Beijing 102413, P.~R.~China}
\newcommand{\cns}{Center for Nuclear Study, Graduate School of Science, University of Tokyo, 7-3-1 Hongo, Bunkyo, Tokyo 113-0033, Japan}
\newcommand{\colorado}{University of Colorado, Boulder, Colorado 80309, USA}
\newcommand{\columbia}{Columbia University, New York, New York 10027 and Nevis Laboratories, Irvington, New York 10533, USA}
\newcommand{\czechtech}{Czech Technical University, Zikova 4, 166 36 Prague 6, Czech Republic}
\newcommand{\dapnia}{Dapnia, CEA Saclay, F-91191, Gif-sur-Yvette, France}
\newcommand{\debrecen}{Debrecen University, H-4010 Debrecen, Egyetem t{\'e}r 1, Hungary}
\newcommand{\elte}{ELTE, E{\"o}tv{\"o}s Lor{\'a}nd University, H - 1117 Budapest, P{\'a}zm{\'a}ny P. s. 1/A, Hungary}
\newcommand{\ewha}{Ewha Womans University, Seoul 120-750, Korea}
\newcommand{\fit}{Florida Institute of Technology, Melbourne, Florida 32901, USA}
\newcommand{\fsu}{Florida State University, Tallahassee, Florida 32306, USA}
\newcommand{\gsu}{Georgia State University, Atlanta, Georgia 30303, USA}
\newcommand{\hanyang}{Hanyang University, Seoul 133-792, Korea}
\newcommand{\hiroshima}{Hiroshima University, Kagamiyama, Higashi-Hiroshima 739-8526, Japan}
\newcommand{\ihepprot}{IHEP Protvino, State Research Center of Russian Federation, Institute for High Energy Physics, Protvino, 142281, Russia}
\newcommand{\illuiuc}{University of Illinois at Urbana-Champaign, Urbana, Illinois 61801, USA}
\newcommand{\inrras}{Institute for Nuclear Research of the Russian Academy of Sciences, prospekt 60-letiya Oktyabrya 7a, Moscow 117312, Russia}
\newcommand{\instpasczech}{Institute of Physics, Academy of Sciences of the Czech Republic, Na Slovance 2, 182 21 Prague 8, Czech Republic}
\newcommand{\isu}{Iowa State University, Ames, Iowa 50011, USA}
\newcommand{\jaea}{Advanced Science Research Center, Japan Atomic Energy Agency, 2-4 Shirakata Shirane, Tokai-mura, Naka-gun, Ibaraki-ken 319-1195, Japan}
\newcommand{\jinrdubna}{Joint Institute for Nuclear Research, 141980 Dubna, Moscow Region, Russia}
\newcommand{\jyvaskyla}{Helsinki Institute of Physics and University of Jyv{\"a}skyl{\"a}, P.O.Box 35, FI-40014 Jyv{\"a}skyl{\"a}, Finland}
\newcommand{\kek}{KEK, High Energy Accelerator Research Organization, Tsukuba, Ibaraki 305-0801, Japan}
\newcommand{\korea}{Korea University, Seoul, 136-701, Korea}
\newcommand{\kurchatov}{Russian Research Center ``Kurchatov Institute", Moscow, 123098 Russia}
\newcommand{\kyoto}{Kyoto University, Kyoto 606-8502, Japan}
\newcommand{\labllr}{Laboratoire Leprince-Ringuet, Ecole Polytechnique, CNRS-IN2P3, Route de Saclay, F-91128, Palaiseau, France}
\newcommand{\lahorelums}{Physics Department, Lahore University of Management Sciences, Lahore 54792, Pakistan}
\newcommand{\lawllnl}{Lawrence Livermore National Laboratory, Livermore, California 94550, USA}
\newcommand{\losalamos}{Los Alamos National Laboratory, Los Alamos, New Mexico 87545, USA}
\newcommand{\lpc}{LPC, Universit{\'e} Blaise Pascal, CNRS-IN2P3, Clermont-Fd, 63177 Aubiere Cedex, France}
\newcommand{\lund}{Department of Physics, Lund University, Box 118, SE-221 00 Lund, Sweden}
\newcommand{\maryland}{University of Maryland, College Park, Maryland 20742, USA}
\newcommand{\mass}{Department of Physics, University of Massachusetts, Amherst, Massachusetts 01003-9337, USA }
\newcommand{\michigan}{Department of Physics, University of Michigan, Ann Arbor, Michigan 48109-1040, USA}
\newcommand{\muenster}{Institut fur Kernphysik, University of Muenster, D-48149 Muenster, Germany}
\newcommand{\muhlenberg}{Muhlenberg College, Allentown, Pennsylvania 18104-5586, USA}
\newcommand{\myongji}{Myongji University, Yongin, Kyonggido 449-728, Korea}
\newcommand{\nagasaki}{Nagasaki Institute of Applied Science, Nagasaki-shi, Nagasaki 851-0193, Japan}
\newcommand{\natmephi}{National Research Nuclear University, MEPhI, Moscow Engineering Physics Institute, Moscow, 115409, Russia}
\newcommand{\newmex}{University of New Mexico, Albuquerque, New Mexico 87131, USA }
\newcommand{\nmsu}{New Mexico State University, Las Cruces, New Mexico 88003, USA}
\newcommand{\ohio}{Department of Physics and Astronomy, Ohio University, Athens, Ohio 45701, USA}
\newcommand{\ornl}{Oak Ridge National Laboratory, Oak Ridge, Tennessee 37831, USA}
\newcommand{\orsay}{IPN-Orsay, Universite Paris Sud, CNRS-IN2P3, BP1, F-91406, Orsay, France}
\newcommand{\peking}{Peking University, Beijing 100871, P.~R.~China}
\newcommand{\pnpi}{PNPI, Petersburg Nuclear Physics Institute, Gatchina, Leningrad region, 188300, Russia}
\newcommand{\riken}{RIKEN Nishina Center for Accelerator-Based Science, Wako, Saitama 351-0198, Japan}
\newcommand{\rikjrbrc}{RIKEN BNL Research Center, Brookhaven National Laboratory, Upton, New York 11973-5000, USA}
\newcommand{\rikkyo}{Physics Department, Rikkyo University, 3-34-1 Nishi-Ikebukuro, Toshima, Tokyo 171-8501, Japan}
\newcommand{\saispbstu}{Saint Petersburg State Polytechnic University, St. Petersburg, 195251 Russia}
\newcommand{\saopaulo}{Universidade de S{\~a}o Paulo, Instituto de F\'{\i}sica, Caixa Postal 66318, S{\~a}o Paulo CEP05315-970, Brazil}
\newcommand{\seoulnat}{Department of Physics and Astronomy, Seoul National University, Seoul 151-742, Korea}
\newcommand{\stonybrkc}{Chemistry Department, Stony Brook University, SUNY, Stony Brook, New York 11794-3400, USA}
\newcommand{\stonycrkp}{Department of Physics and Astronomy, Stony Brook University, SUNY, Stony Brook, New York 11794-3800,, USA}
\newcommand{\tenn}{University of Tennessee, Knoxville, Tennessee 37996, USA}
\newcommand{\titech}{Department of Physics, Tokyo Institute of Technology, Oh-okayama, Meguro, Tokyo 152-8551, Japan}
\newcommand{\tsukuba}{Institute of Physics, University of Tsukuba, Tsukuba, Ibaraki 305, Japan}
\newcommand{\vandy}{Vanderbilt University, Nashville, Tennessee 37235, USA}
\newcommand{\waseda}{Waseda University, Advanced Research Institute for Science and Engineering, 17 Kikui-cho, Shinjuku-ku, Tokyo 162-0044, Japan}
\newcommand{\weizmann}{Weizmann Institute, Rehovot 76100, Israel}
\newcommand{\wigner}{Institute for Particle and Nuclear Physics, Wigner Research Centre for Physics, Hungarian Academy of Sciences (Wigner RCP, RMKI) H-1525 Budapest 114, POBox 49, Budapest, Hungary}
\newcommand{\yonsei}{Yonsei University, IPAP, Seoul 120-749, Korea}
\newcommand{\zagreb}{University of Zagreb, Faculty of Science, Department of Physics, Bijeni\v{c}ka 32, HR-10002 Zagreb, Croatia}
\affiliation{\abilene}
\affiliation{\augie} 
\affiliation{\banaras}
\affiliation{\barc}
\affiliation{\baruch}
\affiliation{\bnlcoll}
\affiliation{\bnlphys}
\affiliation{\caucr}
\affiliation{\charlesczech}
\affiliation{\chonbuk}
\affiliation{\ciae}
\affiliation{\cns}
\affiliation{\colorado}
\affiliation{\columbia}
\affiliation{\czechtech}
\affiliation{\dapnia}
\affiliation{\debrecen}
\affiliation{\elte}
\affiliation{\ewha}
\affiliation{\fit}
\affiliation{\fsu}
\affiliation{\gsu}
\affiliation{\hanyang}
\affiliation{\hiroshima}
\affiliation{\ihepprot}
\affiliation{\illuiuc}
\affiliation{\inrras}
\affiliation{\instpasczech}
\affiliation{\isu}
\affiliation{\jaea}
\affiliation{\jinrdubna}
\affiliation{\jyvaskyla}
\affiliation{\kek}
\affiliation{\korea}
\affiliation{\kurchatov}
\affiliation{\kyoto}
\affiliation{\labllr}
\affiliation{\lahorelums}
\affiliation{\lawllnl}
\affiliation{\losalamos}
\affiliation{\lpc}
\affiliation{\lund}
\affiliation{\maryland}
\affiliation{\mass}
\affiliation{\michigan}
\affiliation{\muenster}
\affiliation{\muhlenberg}
\affiliation{\myongji}
\affiliation{\nagasaki}
\affiliation{\natmephi}
\affiliation{\newmex}
\affiliation{\nmsu}
\affiliation{\ohio}
\affiliation{\ornl}
\affiliation{\orsay}
\affiliation{\peking}
\affiliation{\pnpi}
\affiliation{\riken}
\affiliation{\rikjrbrc}
\affiliation{\rikkyo}
\affiliation{\saispbstu}
\affiliation{\saopaulo}
\affiliation{\seoulnat}
\affiliation{\stonybrkc}
\affiliation{\stonycrkp}
\affiliation{\tenn}
\affiliation{\titech}
\affiliation{\tsukuba}
\affiliation{\vandy}
\affiliation{\waseda}
\affiliation{\weizmann}
\affiliation{\wigner}
\affiliation{\yonsei}
\affiliation{\zagreb}
\author{A.~Adare} \affiliation{\colorado}
\author{S.~Afanasiev} \affiliation{\jinrdubna}
\author{C.~Aidala} \affiliation{\losalamos} \affiliation{\mass} \affiliation{\michigan}
\author{N.N.~Ajitanand} \affiliation{\stonybrkc}
\author{Y.~Akiba} \affiliation{\riken} \affiliation{\rikjrbrc}
\author{R.~Akimoto} \affiliation{\cns}
\author{H.~Al-Bataineh} \affiliation{\nmsu}
\author{H.~Al-Ta'ani} \affiliation{\nmsu}
\author{J.~Alexander} \affiliation{\stonybrkc}
\author{A.~Angerami} \affiliation{\columbia}
% ...
\collaboration{PHENIX Collaboration} \noaffiliation

\date{\today}


\begin{abstract}


The PHENIX experiment at RHIC has measured the centrality dependence of 
the direct photon yield from \AuAu collisions at 
$\sqrt{s_{_{NN}}}=200$~GeV down to $\pt =0.4$~GeV/$c$. Photons are detected 
via photon conversions to $e^+e^-$ pairs and an improved technique is 
applied that minimizes the systematic uncertainties that usually limit 
direct photon measurements, in particular at low $p_T$.  We find an excess 
of direct photons above the $N_{\rm coll}$-scaled yield measured in 
\pp collisions.  This excess yield is well described by an 
exponential distribution with an inverse slope of about 240~MeV/$c$ in the 
\pt range from 0.6--2.0~GeV/$c$. While the shape of the \pt distribution 
is independent of centrality within the experimental uncertainties, the 
yield increases rapidly with increasing centrality, scaling approximately 
with $N_{\rm part}^\alpha$, where 
$\alpha=1.38{\pm}0.03({\rm stat}){\pm}0.07({\rm syst})$.

\end{abstract}

	
\maketitle

\section{Introduction}

Photons are an excellent probe of the hot and dense, strongly 
interacting matter produced in heavy ion collisions~\cite{shuryak}. 
They do not participate in the strong interaction and thus exit the 
system carrying information from the time of their emission, 
allowing a glimpse at the time-evolution of the matter. 
Experimentally we measure a time-integrated history of the 
emission. Photons from hadron decays need to be removed to reveal 
the so-called direct contribution, i.e. photons that are produced 
before the formation of the matter as well as from the matter 
itself. Further removal of the early component, usually considered 
prompt production from $2\rightarrow 2$ scattering of the partons 
from the incoming nuclei, gives access to the radiation emitted 
from the matter. If the matter is in local equilibrium the photon 
spectrum is a time-integrated image of the evolution of the 
temperature and collective motion of the matter as it expands and 
cools.

PHENIX discovered evidence of thermal photons from \AuAu collisions 
at the Relativistic Heavy Ion Collider (RHIC)~\cite{ppg086} ; 
similar findings have recently been reported by ALICE from \PbPb 
collisions at the Large Hadron Collider~\cite{alicephotons}. 
Photons in both energy regimes exhibit a large yield and an 
azimuthal anisotropy~\cite{ppg126,aliceflowproceed} with respect to 
the reaction plane, often referred to as elliptic flow and 
quantified as $v_2$.  Comparing the measured \pt spectra to model 
calculations of thermal photons based on a hydrodynamic evolution 
of the system, microscopic transport models, or a more schematic 
time evolution gives reasonable agreement when assuming an initial 
temperature of 300~MeV or 
above~\cite{dEnterria,Turbide04,huovinen02,srivastava01,alam,Liu08,Zahed,Renk} 
for $\sqsn=200$~GeV $\AuAu$ collisions at RHIC.  However, it is a 
challenge for these types of models to simultaneously explain the 
large observed azimuthal anisotropy of the radiation and the large 
yield~\cite{Eskola,Chaudhari,vHess,Renk,Liu:2012ax}.

The challenge for these model calculations results from the 
interplay between the time evolution of the collective motion and 
the cooling of the matter that emits photons.  In the model 
calculations, the collective motion builds up over time. The flow 
velocity is initially small and increases throughout the collision 
as the matter continues to expand.  The yield of thermal photons is 
expected to be largest early in the collision when the matter is 
the hottest. Theoretical models that create large photon $v_2$ 
typically underestimate the direct photon yield.  Attempts to 
improve hydrodynamic models by implementing next-to-leading-order 
thermal rates~\cite{Teaney}, initial state 
fluctuations~\cite{Renk}, formation time effects~\cite{Liu:2012ax}, 
increased radial flow and enhanced coupling at $T_C$~\cite{vHess}, 
fail to reconcile yield and anisotropy.

To resolve this puzzle, new production mechanisms have been proposed. Some 
enhance the thermal yield in the presence of the strong magnetic field 
perpendicular to the reaction plane, which creates a large 
anisotropy~\cite{Skokov,Muller:2013ila}. Other new mechanisms, such as 
synchrotron radiation~\cite{Goloviznin:2012dy} at the plasma boundary or 
photon production in a glasma phase~\cite{Chiu:2012ij}, create an anisotropy 
due to the initial geometry of the overlap region.

In this paper we present the first measurement of low momentum real direct 
photons from \AuAu collisions at 200 GeV center of mass energy. 
This measurement compliments earlier measurements of direct photons that 
were obtained by extrapolating low mass virtual photons to the real photon 
point \cite{ppg086}. We are able to extend the \pt range down to  
 0.4 GeV/c and provide new information on the centrality dependence 
of the direct photon yield. In particular, the centrality dependence holds 
the promise to help to distinguish between different production mechanisms 
~\cite{Cerny}. 

\section{Experiment}

To measure direct photons, we analyzed large data samples of 
$1.4{\times}10^9$ and $2.6{\times}10^9$ minimum-bias Au$+$Au collisions 
recorded with the PHENIX central arm spectrometers during the 2007 and 
2010 runs, respectively. The main PHENIX detector is described in detail 
elsewhere~\cite{mainNIM}. In addition, a Hadron Blind Detector 
(HBD)~\cite{hbdNIM} was installed, except for part of the 2007 RHIC run 
when only one half of the HBD was installed. 
The data were taken with a 
special field configuration which essentially cancels the magnetic field 
around the beam axis out to about $50\,{\rm cm}$.

Minimum-bias events were triggered using the beam-beam counters (BBC) that 
cover the rapidity region $3.1 < |\eta| < 3.9$ and $2\pi$ in azimuth in 
both beam directions. The BBC information is used to limit the vertex in 
beam direction to $\pm10\,{\rm cm}$ around the nominal position. The 
charge measured in the BBC is used to categorize the event centrality. The 
sample is divided into four centrality classes, 0\%--20\% for the most 
central selection, 20\%--40\%, 40\%--60\% and 60\%--92\% for the most peripheral 
sample.

The raw inclusive photon yield \Nincl is measured through photon 
conversions to \ee pairs in the detector material, which allows us to 
avoid hadron contamination and measure photons down to 
$\ptee=0.4\,{\rm GeV}/c$.  Trajectories and momenta of $e^+$ and 
$e^-$ are determined using the drift chambers and the pad chambers  
that measure the deflection in the axial magnetic field together with the
interaction vertex location.  We require a 
minimum \pt of $200\,{\rm MeV}/c$. The energy is determined with the 
electromagnetic calorimeter (EMCal). The $e^+$ and $e^-$ are identified 
utilizing the ring-imaging \v{C}erenkov detector by requiring a 
minimum of three phototubes associated with both charged tracks at the 
expected ring radius as well as requiring the respective energy/momentum 
ratios to be greater than 0.6.

\section{Data Analysis}

\begin{figure}[htb] 
  \includegraphics[width=1.0\linewidth]{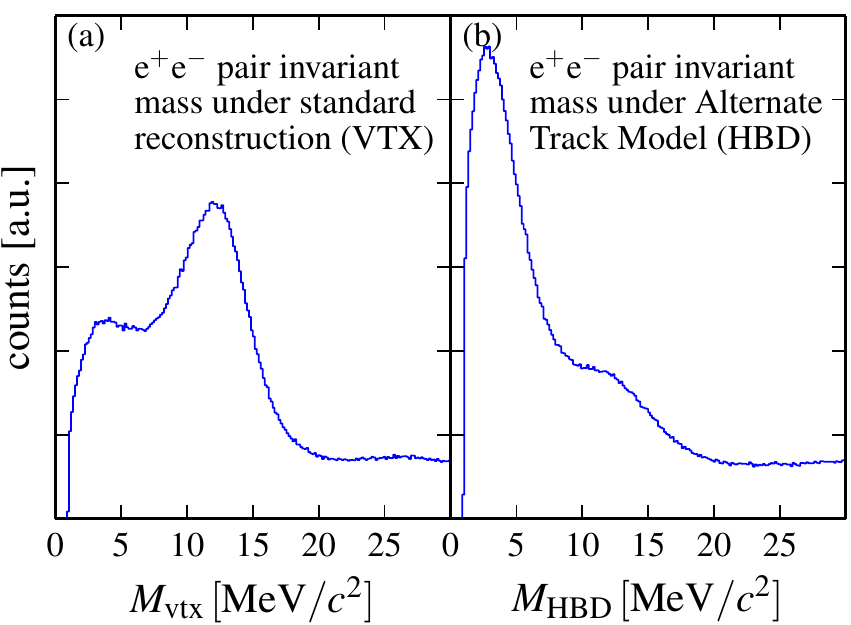}
  \caption{
Histograms of the $e^+$$e^-$ pair invariant-mass distribution from data.  
Panel (a) shows the distribution of masses calculated with the normal 
reconstruction algorithm (vtx).  Panel (b) shows the distribution of 
masses calculated with the alternate track model assumption (HBD).
  }
  \label{fig:1dmassplot}
\end{figure}

We select photons that 
converted in the readout plane of the HBD that is located at a 
radius of $60\,{\rm cm}$ and has a radiation length $X/X_0 \approx 
2.5\%$. Our method to identify photon conversions uses only 
the PHENIX central arm detectors, with the HBD playing no active role.   
Because the draft chambers are located at $\approx$220~cm radially from 
the beam axis, the momentum reconstruction algorithm needs to assume 
where the particles originate. In the standard algorithm all charged 
particles are reconstructed as if they came from the event vertex. 
This procedure mismeasures the momentum vector for $e^+$ and
$e^-$ from photon conversions in the HBD. For conversions in the HBD readout plane the artificial opening 
angle of the \ee pair is $\approx$10 mrad and the pair momentum increases 
by 1\%--2\%. As a result the \ee pair is reconstructed with an average 
mass of  $M_{\rm vtx} \approx 12\,{\rm MeV}/c^2$, as is shown in the 
invariant-pair-mass distribution of Fig.~\ref{fig:1dmassplot}a. 
The first  peak in the mass plot at a few MeV/$c^2$ is from $\pi^0$ 
Dalitz decays, along with a small number of pairs from photon conversions 
before the HBD readout plane.

\begin{figure}[htb]
  \includegraphics[width=1.0\linewidth]{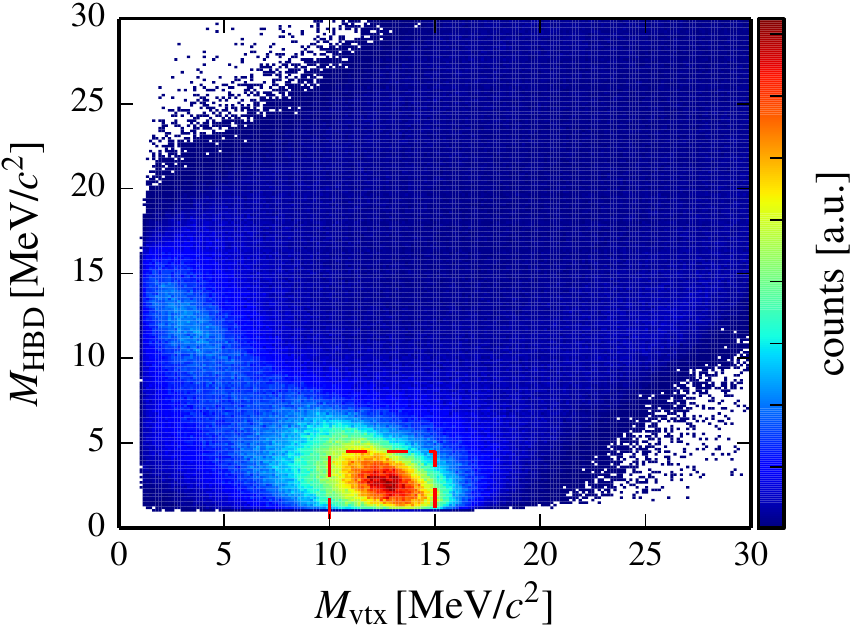}
  \caption{
(Color online) A view of the cut space used for the conversion photon identification.  
The mass as calculated under the standard reconstruction algorithm (vtx) 
is plotted on the horizontal axis, while the mass as calculated under the 
alternate track model (HBD) is plotted on the vertical axis. The 
dotted (red) box indicates the region used to identify photon conversions.
  }
  \label{fig:2dmassplot}
\end{figure}

The momenta of all low mass \ee pairs are recalculated assuming 
that they originated at the HBD readout plane. If the \ee pair is indeed 
a conversion pair from the readout plane, the relative momentum resolution of 
the pair is approximately 
$\sigma_{p_T}^{ee}/p_T^{ee}=0.9\%{\oplus}0.5\%p_T^{ee}$ and the \ee pair mass  recalculated with the HBD back plane as origin ($M_\mathrm{HBD}$)
is a few MeV/$c^2$, 
consistent with the experimental resolution. For all other \ee pairs, 
the momentum vectors are now mismeasured, in particular \ee pairs from 
$\pi^0$ Dalitz decays are now reconstructed with larger opening angles and thus 
shifted upward in \ee pair mass. The recalculated mass spectrum is 
shown in panel (b) of Fig.~\ref{fig:1dmassplot}. Plotting the yield as a
function of $M_\mathrm{HBD}$ versus the mass calculated with the vertex as origin 
($M_{\rm vtx}$), as shown in Fig.~\ref{fig:2dmassplot}, allows one to 
clearly isolate the conversions in the HBD readout plane. We select photon 
conversions by a two dimensional cut $10 < M_{\rm vtx} < 15\ {\rm MeV}/c^2$ 
and $M_\mathrm{HBD} < 4.5\ {\rm MeV}/c^2$, illustrated by the red 
dashed box.  Note that the large distance from the true event 
vertex and the relatively thick HBD readout plane (in terms of radiation 
length $X_0$) with no comparable radiating material nearby makes  
identification of the converted photons very accurate: a full {\sc geant} Monte 
Carlo simulation \cite{brun1993geant} shows that the purity of this sample is 99\%, with most 
of the remaining 1\% being photon conversions at other radii.

A subset of this inclusive conversion photon sample \Nincl is tagged 
statistically as photons from $\pi^0$ decays if they reconstruct the 
$\pi^0$ mass with a second, photon-like shower taken from the EMCal. Note 
that this is done in bins of \ptee, the transverse momentum of the 
converted photons, not in bins of $\pi^0$ \pt.  A cut on the shower shape 
of this second EMCal shower is used to remove most hadrons. False tagging 
from hadron showers in the EMCal is further reduced by applying a lower 
threshold on the cluster energy. For the 2010 data we applied an 
$E_\mathrm{clus}>$0.4~GeV cut, which is just above the EMCal response for 
minimum ionizing particles. For the 2007 data, a higher threshold of 
0.6~GeV was necessary due to a cut on the shower energy that was 
introduced during data production.

\begin{figure}[htb]
  \includegraphics[width=1.0\linewidth]{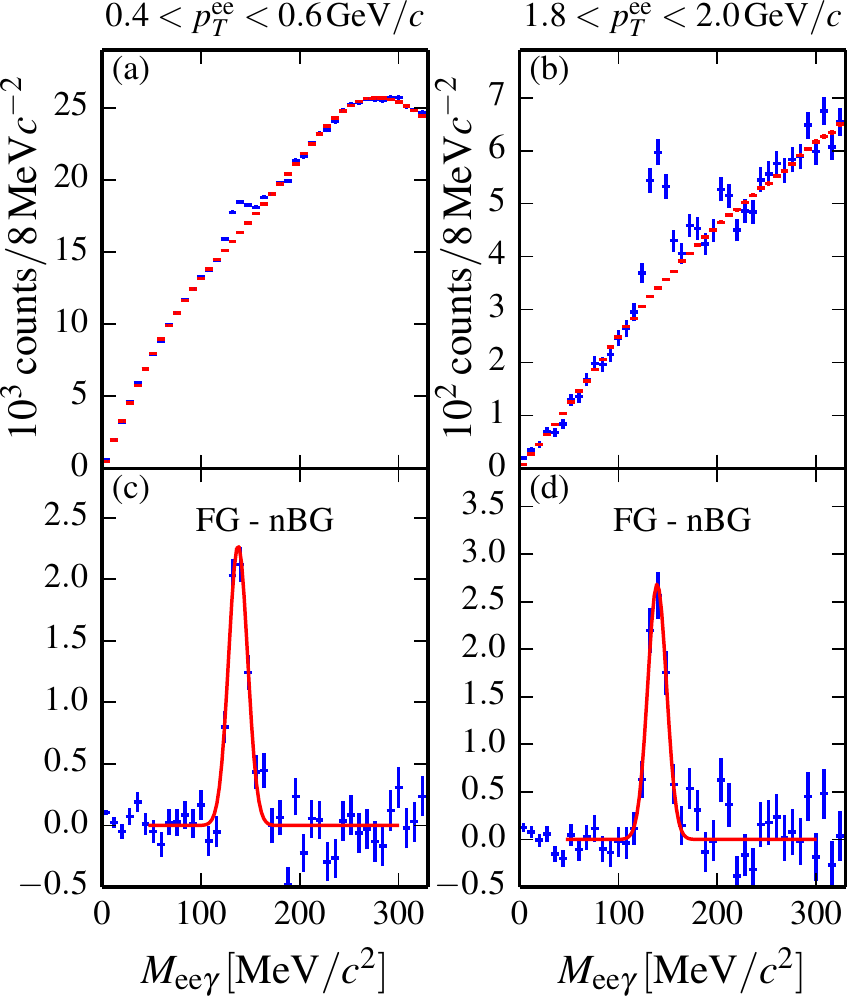}
  \caption{(Color online) Histograms of the $e^+e^-\gamma$ 
invariant-mass distributions 
for two different \ptee bins.  The left column (a),(c) displays the mass 
for $0.4 < \ptee < 0.6$ GeV/$c$, the right column (b),(d) displays the 
mass for $1.8 < \ptee < 2.0$ GeV/$c$.  The top row (a),(b) shows the 
$e^+e^-\gamma$ invariant-mass foreground distribution in blue, with the 
normalized background 
distribution from the mixed events in red.  The bottom row (c),(d) shows 
the isolated pion peak after subtraction of the normalized background.  
The masses are calculated from the HBD readout plane origin assumption on 
the electron tracks.  The centrality bin is 0\%--20\%.
  }
  \label{fig:pi0massplot}
\end{figure} 

\subsection{Relative Photon Yield}

In each \ptee bin the number of $\pi^0$ tagged photons (\Npi) is 
determined by integrating the $\ee \gamma$ mass distribution around the 
$\pi^0$ mass after subtraction of the mixed-event combinatorial 
background.   Figure~\ref{fig:pi0massplot} shows the mass distributions 
before and after subtracting the mixed-event background for two example 
\ptee bins (0.4--0.6~GeV/$c$ and 1.8--2.0~GeV/$c$) for central collisions 
(0\%--20\%), which have the smallest signal-to-background ratio. The $\pi^0$ 
peak extraction method has less than 4\% systematic uncertainty on the 
$\pi^0$ tagged photon yield, which is assumed to be independent between 
neighboring \ptee bins and thus folded into the statistical uncertainties.

In a given \ptee bin the true number of inclusive photons 
$\gamma^{\rm incl}$ and photons from $\pi^0$ decays $\gamma^{\pi^0}$ 
are related to the measured quantities \Nincl and \Npi as follows:
\begin{equation}
	\Nincl = \varepsilon_{ee}a_{ee} \ c \gamma^{\rm incl} \
	\label{eqn:incl},
\end{equation}
\begin{equation}
	\Npi = \varepsilon_{ee}a_{ee} \ c \langle\varepsilon_\gamma f\rangle \gamma^{\pi^0}.
	\label{eqn:pitag},
\end{equation}
where $c$ is the probability that the photon converts in the HBD readout 
plane, $\varepsilon_{ee}$ is the reconstruction efficiency of the \ee pair 
and $a_{ee}$ is the factor describing that both $e^+$ and $e^-$ are in the 
detector acceptance. The factor $f$ is the conditional acceptance that 
after one photon from a $\pi^0$ decay was reconstructed as \ee conversion 
pair, the partner photon falls into the acceptance of the EMCal. The 
probability that the partner photon is reconstructed is given as 
$\varepsilon_\gamma$. The product $\varepsilon_\gamma f$ is averaged over 
all possible \pt of the partner photon, indicated by $\langle 
\varepsilon_\gamma f\rangle$.

Because \Nincl and \Npi are both measured in terms of the \ptee of the 
converted photon, the efficiency and acceptance factors for the \ee pair 
as well as the conversion probability explicitly cancel in the ratio 
$\Nincl/\Npi$. This ratio can be converted into \Rgamma, the ratio of the 
yield of true inclusive photons $\gamma^{\rm incl}$ to the yield of 
true photons from hadron decays $\gamma^{\rm hadron}$:

\begin{equation}
	R_\gamma = \frac{\gamma^{\rm incl}}{\gamma^{\rm hadron}} = \frac{\langle\varepsilon_\gamma f \rangle \left(\frac{N^{\rm incl}_\gamma}{N^{\pi^0,{\rm tag}}_\gamma}\right)_{\rm Data}}{\left(\frac{\gamma^{\rm hadron}}{\gamma^{\pi^0}}\right)_{\rm Sim}}
	\label{eqn:rgamma}
\end{equation}

All terms in Eq.~\ref{eqn:rgamma} are a function of converted photon 
\ptee. \Rgamma will be unity for a given \ptee bin if all photons result 
from hadron decays, or larger than unity if direct photons are present in 
the sample.  The excess above unity is a measure of the direct photon 
content in the bin. In the following we discuss all terms in detail.

\begin{figure}[htb]
  \includegraphics[width=1.0\linewidth]{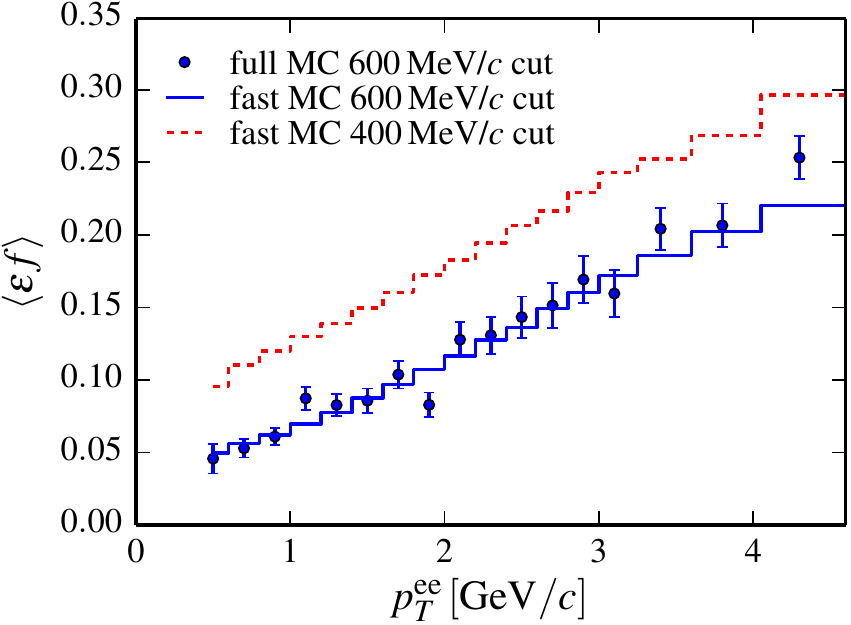}
  \caption{ (Color online)
Average conditional acceptance \ef to detect a photon from a $\pi^0$ decay in the 
EMCal, if the other photon converted in the HBD electronics and was reconstructed
as an \ee pair. The abscissa gives \ptee, the \pt of the \ee pair. The \pt cut
of 0.6 (2007) and 0.4 GeV/$c$ (2004) is on the photon detected in the EMCal. 
For the \pt cut of 0.6 GeV/$c$ we show the results for 
two methods, a full MC simulation (points) and a fast MC simulation (histogram). 
For the \pt cut of 0.4 GeV, the 
fast MC simulation is shown as dashed histogram.  
}
  \label{fig:ef} 
\end{figure} 

\begin{table}[htb]
\caption{
Summary of systematic uncertainties on \Rgamma. The $\pi^0$ reconstruction 
uncertainty is uncorrelated between data points (type A); type B 
uncertainties are \pt-correlated, and type C are uncertainties that can change  
$R_\gamma$ for all \pt by a constant multiplicative factor.  
}
\begin{ruledtabular} \begin{tabular}{lccc}
Source                          & $\sigma_{\rm syst}/R_\gamma$ & Type & \\
\hline
$\pi^0$ reconstruction	 	&     &   & \\
(tagged photon yield)	  	& 4\% & A & \\
$\gamma$ purity 	  	& 1\% & C & \\
\\
conditional acceptance $\langle\varepsilon f\rangle$ & & & \\
energy scale 		 	& 4\% & B & \\
conversion loss 	 	& 2\% & C & \\
$\gamma$ efficiency 	 	& 1\% & B & \\
active area 		 	& 1\% & C & \\
input \pt spectra       	& 1\% & B & \\
\\
$\gamma^{\rm hadron}/\gamma^{\pi^0}$  & & & \\
$\eta/\pi^0$ ratio 	 	& 2.2\%  & C & \\
other mesons			& $<$1\% & C & \\
\end{tabular} \end{ruledtabular}
\label{tab:uncertainties}
\end{table} 

\begin{figure}[hb]
  \includegraphics[width=0.998\linewidth]{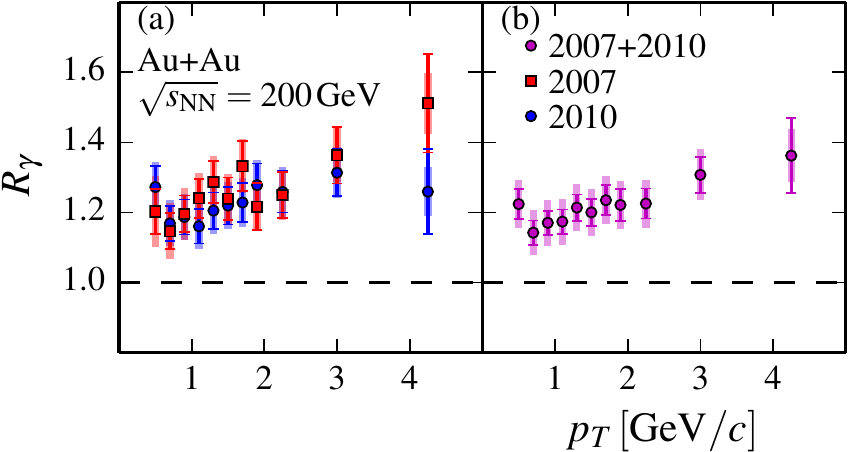}
  \caption{ (Color online)
\emph{(a)} Ratio \Rgamma as function of photon \pt from the 2007 
(red open square) and from the 2010 
data sets (blue closed circle) in minimum-bias \AuAu collisions. 
Statistical uncertainties are dominated by the $\pi^0$ yield extraction. 
They are plotted as vertical lines. All other systematic uncertainties are 
added in quadrature and shown as filled boxes. \emph{(b)} \Rgamma in the 
combined 2007+2010 measurement.
}
  \label{fig:rg_mb} 
\end{figure}

The numerator of Eq.~\ref{eqn:rgamma} includes the measured ratio 
$\Nincl/\Npi$, and the efficiency and acceptance correction for pion 
tagging, \ef.  Figure~\ref{fig:ef} shows 
\ef for the min. bias data sets of 2007 and 2010.  It increases 
monotonically with \ptee and is lower for the larger \pt cut on the second photon. 
These trends can be understood in terms of decay kinematics and average \pt of 
the tagged $\pi^0$. At higher \ptee the average \pt of the tagged
$\pi^0$ is larger, the opening angle between the decay photons becomes smaller and
the probability to have both decay photon in the PHENIX acceptance increases.
Consequently \ef increases with \ptee. A larger \pt-cut on the second photon 
increases the minimum $\pi^0$ \pt necessary for both photons to be accepted 
at a given \ptee, thus \ef is larger for the lower \pt cut. The ratio of \ef 
for the two different \pt cuts is as large as a factor of 2 at the lowest 
\ptee and decreases towards higher \ptee. Because the final result for \Rgamma is 
proportional to \ef, varying the \pt cut provides a powerful
cross check for the measurement. 

We developed two different methods to determine 
$\langle\varepsilon_\gamma f\rangle$. For the 2007 data 
a {\sc geant} Monte Carlo simulation of the detector response to $\pi^0$ decays 
is performed. In the simulation one photon is forced to convert in the HBD 
readout plane. The simulated $\pi^0$ decays are then embedded into real 
data to account for occupancy effects in the EMCal. The events are 
analyzed through the full reconstruction chain to extract 
$\langle\varepsilon_\gamma f\rangle$. This method is computationally very 
intensive and thus limited by statistical uncertainties. To overcome these 
we developed a fast simulation. It accounts for the detector acceptance 
and variations of the active detector areas with time. The single photon 
response is parametrized based on a {\sc geant} Monte Carlo simulation of 
single photons. To test the fast simulation we compared its result for \ef
in Fig.~\ref{fig:ef} for the 2007 data 
to the one determined with the full {\sc geant} simulation; the two methods 
agree within statistical uncertainties. For the 2010 data we used the fast 
simulation. We also compare results on \Rgamma for  \pt cuts on the 
second photon between 0.3 and 0.6 GeV/$c^2$ and find that the results are 
consistent well within the systematic uncertainties on \ef discussed below. 

The denominator of Eq.~\ref{eqn:rgamma} is the ratio of photons from all 
hadron decays ($\gamma^{\rm hadron}$) to those from $\pi^0$ decays
($\gamma^{\pi^0}$). To evaluate this ratio, the per-event yields
$\gamma^{\pi^0}$ and $\gamma^{\rm hadron}$ are determined using the PHENIX 
meson decay generator {\sc exodus}, which is discussed in detail 
in Ref.~\cite{ppg088}.  

For each centrality class, a fit to the measured per-event yields for
charged and neutral pions \cite{ppg014, ppg026} is used to generate
$\pi^0$'s that then are decayed to photons according to known branching
ratios and decay kinematics based on Ref.~\cite{PDG}. The resulting photon
spectrum is the per-event yield $\gamma^{\pi^0}$ as a function of photon
$p_T$. To generate $\gamma^{\rm hadron}$, the contributions 
from decays of $\eta$, $\omega$, and $\eta^\prime$ are determined 
using the same procedure and then added to $\gamma^{\pi^0}$. The shape of
the \pt spectra for $\eta$, $\omega$, and $\eta^\prime$ are derived from
the $\pi^0$ spectrum by replacing \pt with $m_T = \sqrt{m^2_{\rm hadron} -
m^2_{\pi^0} + \pt^2}$. For $\eta$ and $\omega$ this is consistent with
published data~\cite{ppg055, ppg118}; for $\eta^\prime$ no data are
available. The absolute normalization of the $\eta$ per-event yield is set
using a value of $\eta/\pi^0 = 0.46 \pm 0.06$~\cite{ppg051,ppg133} at 
\pt = 5 GeV/$c$. For the $\omega$ and $\eta^\prime$ the absolute yield is
set to $\omega/\pi^0 = 0.9\pm0.06$ and $\eta^\prime/\pi^0=0.25\pm0.075$,
again at 5 GeV/$c$ (see \cite{ppg088}).

\subsection{Systematic Uncertainties}

Several sources contribute to systematic uncertainties on \ef. The largest 
one is 4\% and accounts for the uncertainties of the energy scale and the 
energy resolution. These translate directly into an uncertainty in the 
number of photons that pass the lower EMCal threshold and thus become 
candidates for $\pi^0$ tagging. The second largest uncertainty (2\%) is on 
the number of photons that are lost because they convert to \ee pairs in 
the detector material in front of the EMCal and are not reconstructed as 
single showers. The active area of the detectors was studied as a function 
of time, and the resulting systematic uncertainty on \ef is smaller than 
1\%. Varying the $\pi^0$ input distribution with the uncertainties on the 
data results in a 1\% uncertainty on \ef. Lastly, the uncertainty on the 
photon reconstruction efficiency is also small (1\%), estimated by varying 
the shower shape cuts, redoing the analysis and recalculating the 
correction, and comparing the results. All other systematic effects were 
found to be negligible.

Systematic uncertainties on $\gamma^{\rm hadron}/\gamma^{\pi^0}$ are dominated 
by the accuracy with which $\eta/\pi^0$ is known. Because the $\pi^0$
contribution to $\gamma^{\rm hadron}$ is $\approx$80\%, the systematic
uncertainty on the $\pi^0$ spectra largely cancels, leaving the 
$\eta/\pi^0$ ratio as the dominant source of 
systematic uncertainties. The uncertainty on \Rgamma also includes 
possible deviations from scaling with $m_T$ and uncertainties on the other 
meson yields. The total uncertainty is less than 2.5\%. All systematic 
uncertainties on \Rgamma are summarized in Table~\ref{tab:uncertainties}.

\begin{figure}[htb]
  \includegraphics[width=1.0\linewidth]{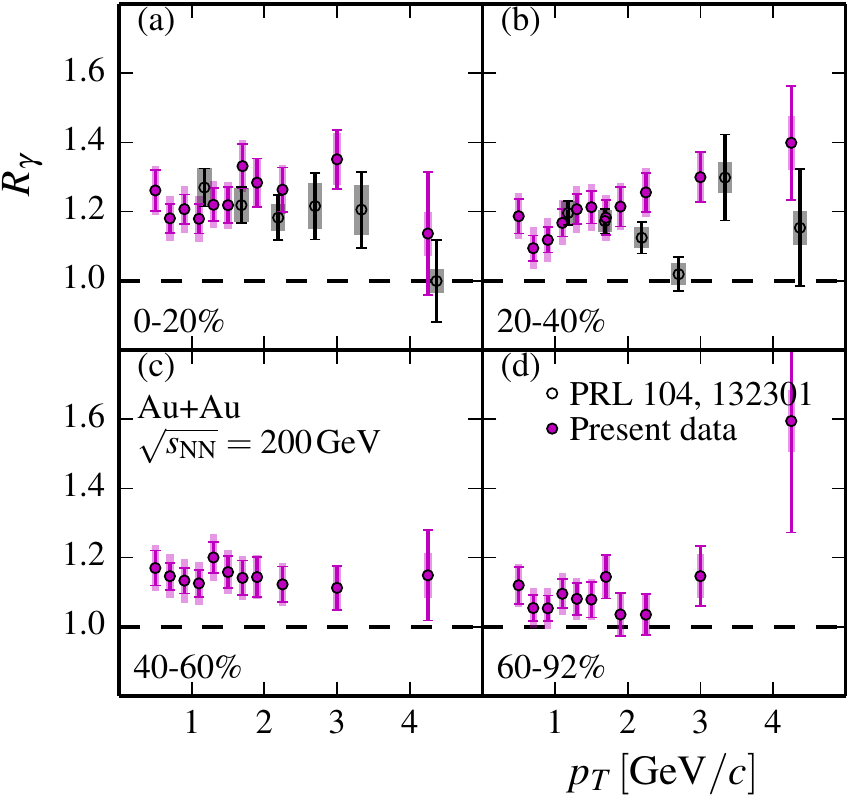}
  \caption{ (Color online)
Ratio $R_\gamma$ as function of photon \pt 
for the combined 2007 and 2010 data sets in centrality 
bins 0\%--20\%, 20\%--40\%, 40\%--60\% and 60\%--92\%. Statistical uncertainties 
plotted as vertical lines are dominated by the $\pi^0$ yield extraction. 
All other systematic uncertainties are added in quadrature and shown as 
filled boxes. On panels (a) and (b) we also show earlier results from 
Ref.~\cite{ppg086}, obtained by extrapolating virtual photons to 
zero mass. 
}
  \label{fig:rg}
\end{figure} 

\section{Results and Discussion}

Figure~\ref{fig:rg_mb} compares our results for \Rgamma in minimum-bias 
collisions from the 2007 and 2010 data sets separately, while 
Figure~\ref{fig:rg} shows the same quantity for the four centrality 
selections. Here we used the full {\sc geant} simulation for the 2007 data, and 
the fast Monte Carlo simulation for the 2010 data. \Rgamma from the two 
data sets agree well within statistical errors.  Figure~\ref{fig:rg}  
also includes data from an earlier publication~\cite{ppg086}, in which 
\Rgamma was obtained by extrapolating virtual photons 
to $m=0$ for the two central bins and $\pt>1.0\,{\rm GeV}/c$. 
The \Rgamma was used to calculate the direct-photon \pt spectra shown 
in~\cite{ppg086}; here we show the corresponding data points.  
We observe no statistically 
significant difference between the \Rgamma measured from real and virtual 
photons.  However, given the uncertainties, we cannot rule out a 
difference of up to 15\%, as is estimated in Ref.~\cite{Zahed}. 
The \Rgamma shows 
a statistically significant excess of photons above those expected from 
hadron decays, and this excess increases with centrality.

\begin{figure}[htb]
  \includegraphics[width=1.0\linewidth]{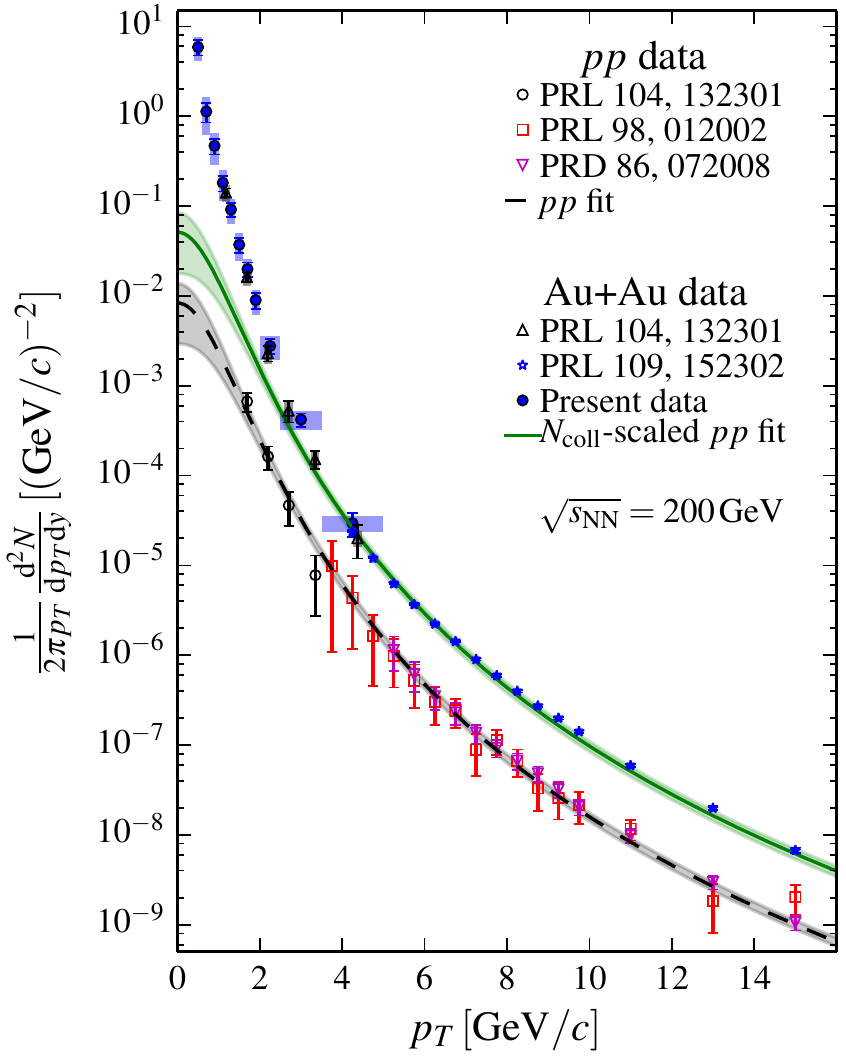}
  \caption{ (Color online)
Direct photon \pt spectra for minimum-bias Au$+$Au collisions from this 
measurement (solid symbols) and Au$+$Au and \pp collisions (open symbols). Open circles 
and up triangles: low \pt spectrum obtained with virtual photons in \pp and 
\AuAu~\cite{ppg086}. Open squares and down triangles: spectrum of real 
photons, measured in the EMCal in \pp. Open squares are 2003 
data~\cite{ppg060}, open down triangles are 2006 data~\cite{ppg136}. Open stars: 
spectrum with real photons, measured in the EMCal in Au$+$Au in 
2004~\cite{ppg139}. The dashed line is a fit to the combined set of \pp 
data, extrapolated below 1\,GeV/$c$, and the solid line the \pp fit 
scaled with the number of minimum-bias Au$+$Au collisions.  Bands around 
lines denote $1\sigma$ uncertainty intervals in the parameterizations of 
the \pp data and the uncertainty in \Ncoll, added in quadrature.
}
  \label{fig:direct_yield_mb}
\end{figure}

\begin{figure}[htb]
  \includegraphics[width=1.0\linewidth]{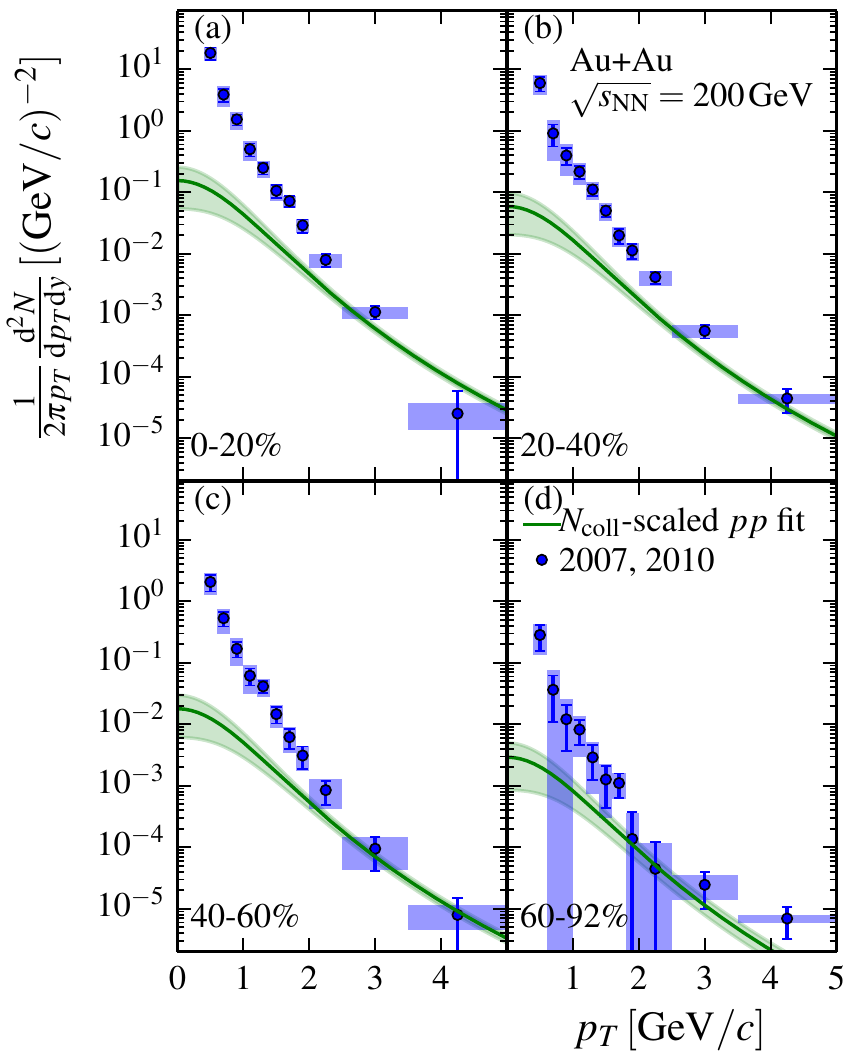}
  \caption{ (Color Online)
Direct photon \pt spectra in centrality bins 0\%--20\%, 20\%--40\%, 40\%--60\% and 
60\%--92\%. Widths of filled boxes indicate bin widths in this analysis.  
The green bands show a \Ncoll-scaled modified power-law fit to the PHENIX 
\pp data and its extrapolation below 1\,GeV/$c$, cf.\ 
Fig.~\protect\ref{fig:direct_yield_mb}. One-sided errors denote $1\sigma$ upper limits, 
other uncertainties are as in Fig.~\protect\ref{fig:direct_yield_mb}.
  }
  \label{fig:pt_direct}
\end{figure} 

To combine the data sets we apply the corrections calculated from the fast 
simulation for both the 2007 and 2010 data (after verifying consistency 
between the corrections calculated for the 2007 data with both the fast 
Monte Carlo and full {\sc geant}) and average the numerators in 
Eq.~\ref{eqn:rgamma} for the 2007 and 2010 data sets. While the correction 
factor \ef is different for the two data sets (due to differences in 
detector dead areas and the different minimum photon energy cuts applied), 
the systematic uncertainties are the same. Next we determine the direct 
photon yield from the combined $R_\gamma$ for each \pt bin:
\begin{equation}
	\gamma^{\rm direct} = (R_\gamma - 1)\gamma^{\rm hadron},
    \label{eqn:directYield}
\end{equation}
were $\gamma^{\rm hadron}$ is the invariant yield of photons from hadron 
decays, which we calculate from measured charged and neutral pion spectra, 
as described above.  At this point a systematic uncertainty of 10\% on the 
shape of the input $\pi^0$ distribution for the generator needs to be 
included~\cite{ppg088} (this mostly cancels in the denominator of 
$\Rgamma$, but no longer cancels in Eq.~\ref{eqn:directYield}). The 
measurement was cross-checked and found consistent with the direct photon 
spectrum calculated using the fully corrected measured inclusive photon 
spectrum~\cite{ppg088} via the relation $\gamma^{\rm direct} = \left(1 - 
1/\Rgamma \right)\gamma^{\rm incl}$, which has much larger systematic 
uncertainties because the conversion probability, the $e^+e^-$ pair 
efficiency and acceptance do not cancel.

Figure~\ref{fig:direct_yield_mb} shows the direct photon \pt spectra for 
minimum bias and our previously published \AuAu data from 
Ref.~\cite{ppg086} and~\cite{ppg139}. Also shown are the \pp photon data from PHENIX. The lowest \pt points (open circles) come from a virtual photon 
measurement~\cite{ppg086}, while the open squares and open triangles are 
from the analysis of the 2003~\cite{ppg060} and 2006~\cite{ppg136} 
data sets, respectively.  The dashed curve is the joint fit to 
the \pp data with a functional form $a\left(1+\frac{\pt^2}{b}\right)^c$.  
This shape was used in Ref.~\cite{ppg086}.  Including new data in 
the fit~\cite{ppg136}, we find parameters $a=(8.3\pm 7.5)\times 10^{-3}$, 
$b=2.26\pm 0.78$ and $c=-3.45\pm 0.08$.  Note that the systematic 
uncertainties are highly correlated.  Also, the lowest actual data point 
in the fit is at \pt=1\,GeV/$c$.

The solid curve in Fig.~\ref{fig:direct_yield_mb} is the \pp fit scaled by 
the corresponding average number of binary collisions, $N_{\rm part}$,
for minimum-bias collisions, as calculated from a Glauber Monte Carlo 
simulation~\cite{annurev2007}.  Below $\pt=3\,{\rm GeV}/c$, an enhancement 
above the expected prompt production (\pp) is observed. The enhancement 
has a significantly smaller inverse slope than the $N_{\rm coll}$ scaled 
\pp contribution.  

Figure~\ref{fig:pt_direct} shows that we observe similar behavior when 
investigating the centrality dependence in more detail. The solid curves 
are again the \pp fit scaled by the respective number of binary 
collisions, and they deviate significantly from the measured yields below 
3\,GeV/$c$.

\begin{figure}[htb]
  \includegraphics[width=1.0\linewidth]{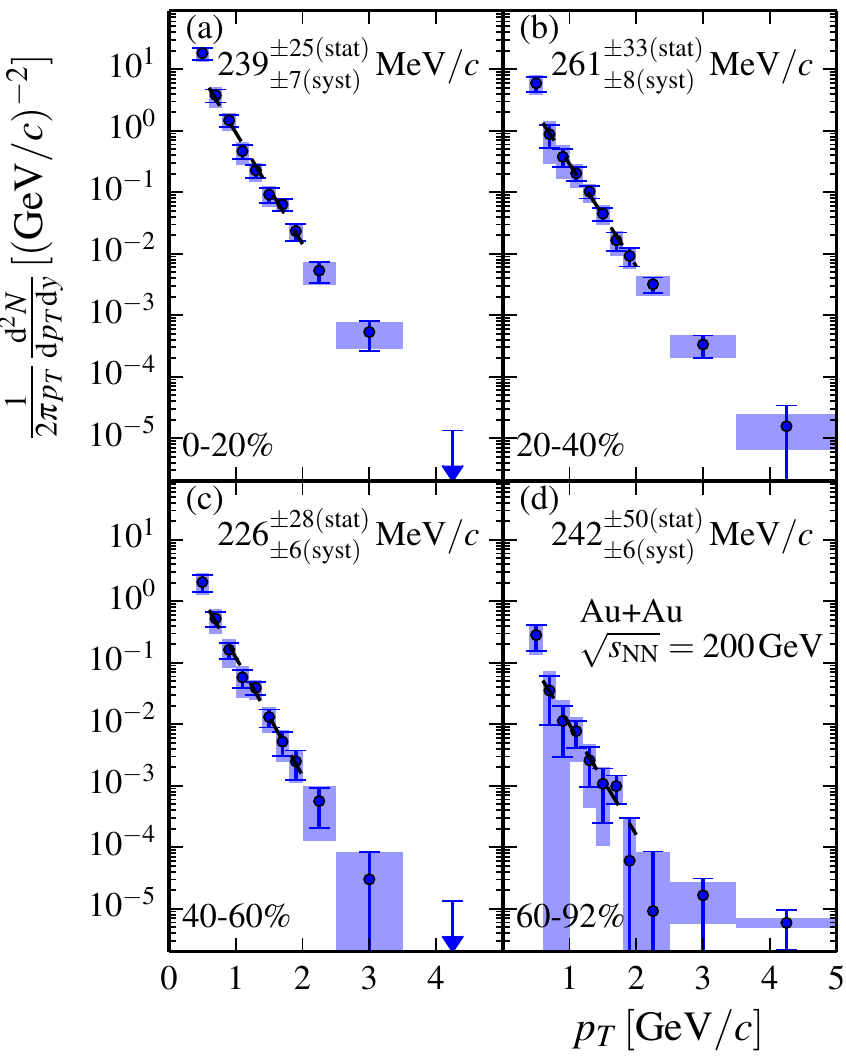}
  \caption{ (Color online)
Direct photon \pt spectra after subtraction of the \Ncoll scaled \pp 
contribution in centrality bins 0\%--20\%, 20\%--40\%, 40\%--60\% and 60\%--92\%. 
Uncertainties are plotted as in Fig.~\ref{fig:pt_direct}. Dashed lines are 
fits to an exponential function in the range $0.6\,{\rm GeV}/c < \pt < 
2.0\,{\rm GeV}/c$. 
}
  \label{fig:pt_thermal}
\end{figure} 

\begingroup 
\squeezetable
\begin{table}[htbp]
\caption{ 
The number of nucleon participants \Npart, number of binary 
nucleon-nucleon collisions \Ncoll, and constituent-quark participants 
\Nqpart vs centrality bin.   Also shown are the values of local inverse 
slopes in the \pt range 0.6 to 2 GeV/$c$ of the direct 
photon spectra, after subtracting the \Ncoll scaled \pp results.
}  
\begin{ruledtabular} \begin{tabular}{rcccl}
Centrality 
& \Ncoll 
& \Npart 
& \Nqpart 
& $T_{\rm eff}$~(MeV/$c$) \\
\hline
0\%--20\%  
& $770.6 \pm 79.9$ 
& $277.5 \pm 6.5$ 
& $735.2 \pm 14.6$ 
& $239 \pm 25\pm 7$ \\
20\%--40\% 
& $282.4 \pm 28.4$ 
& $135.6 \pm 7.0$ 
& $333.2 \pm 10.7$ 
& $260 \pm 33 \pm 8$ \\
40\%--60\% 
& $82.6 \pm 9.3$   
& $56.0 \pm 5.3$  
& $126.6 \pm 6.1$  
& $225 \pm 28 \pm 6$ \\
60\%--92\% 
& $12.1 \pm 3.1$   
& $12.5 \pm 2.6$  
& $25.8  \pm 4.0$  
& $238 \pm 50 \pm 6$ \\
0\%--92\%  
& $251.1 \pm 26.7$ 
& $106.3 \pm 5.0$ 
& $268.8  \pm 8.2$ 
& $242 \pm 28 \pm 7$
\end{tabular} \end{ruledtabular}
\label{tab:data}
\end{table}  
\endgroup

\begin{figure}[htb]
  \includegraphics[width=1.0\linewidth]{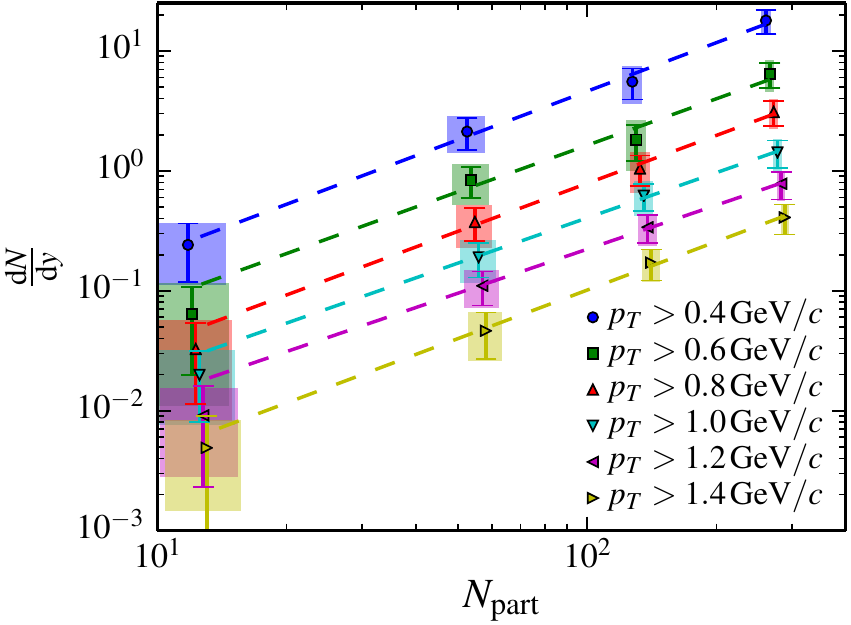}
  \caption{  (Color online)  
Integrated thermal photon yields as a function of \Npart for different 
lower \pt integration limits. The dashed lines are independent fits to a 
power-law.
  }
  \label{fig:int_yield}
\end{figure}

Finally the direct photon contribution from prompt processes (as estimated 
by the $N_{\rm coll}$ scaled \pp\ direct photon yield, shown by the curve in 
Fig.~\ref{fig:pt_direct}) is subtracted to isolate the radiation unique 
to heavy ion collisions.  The results are depicted in 
Fig.~\ref{fig:pt_thermal}.  While the origin of this additional 
radiation cannot be directly established (it could be for instance thermal 
and/or initial state radiation, or the dominant source could even be \pt 
dependent), it is customary to fit this region with an exponential and 
characterize the shape with the inverse slope.  Accordingly, shown on each 
panel is a fit to an exponential function in the range 
$0.6<\pt<2\,{\rm GeV}/c$. The inverse slopes are approximately 
$240\,{\rm MeV}/c$ independent of centrality, see Table~\ref{tab:data}. 
In contrast, the yield clearly increases with centrality. We have 
quantified this by integrating the photon yield above a threshold \ptmin. We varied the threshold from $0.4$ to $1.4\,{\rm GeV}/c$ to show 
that the centrality dependence does not result from a change of shape at low 
\pt (see Fig.~\ref{fig:int_yield}).

\begin{table}[htbp]
  \caption{  
Fitted parameters from fitting power-law fits $\frac{{\rm d} 
N}{{\rm d} y} = A N_{\rm part}^\alpha$ for integrated yields with 
different lower \ptee limits.
  }  
\begin{ruledtabular} \begin{tabular}{ccccc}
&    \ptmin      & & &  \\ 
&    (GeV/$c$)     & $\alpha$ & $A$ &  \\ 
\hline
&    0.4  & $1.36 \pm 0.08 \pm 0.08$ & $(7.85 \pm 2.96 \pm 4.52) \times 10^{-3}$   &  \\
&    0.6  & $1.41 \pm 0.14 \pm 0.12$ & $(2.20 \pm 1.54 \pm 1.64) \times 10^{-3}$   &  \\
&    0.8  & $1.42 \pm 0.07 \pm 0.11$ & $(1.07 \pm 0.39 \pm 0.75) \times 10^{-3}$   &  \\
&    1.0  & $1.35 \pm 0.06 \pm 0.07$ & $(7.70 \pm 2.32 \pm 4.37) \times 10^{-4}$   &  \\
&    1.2  & $1.36 \pm 0.09 \pm 0.07$ & $(3.90 \pm 1.79 \pm 2.81) \times 10^{-4}$   &  \\
&    1.4  & $1.40 \pm 0.06 \pm 0.10$ & $(1.63 \pm 0.47 \pm 1.11) \times 10^{-4}$   &  \\
\end{tabular} \end{ruledtabular}
  \label{tab:powerfit}
\end{table} 

The yield increases with a power-law function $N_{\rm part}^\alpha$; 
this is illustrated by the linear rise of the yield with \Npart in the 
logarithmic representation shown on Fig.~\ref{fig:int_yield} together 
with fits to $A N_{\rm part}^\alpha$. The fit parameters are shown in 
Table~\ref{tab:powerfit}. The same power is observed independent of the 
\pt cutoff, consistent with the spectra having the same shape independent 
of centrality. A simultaneous fit to the data in 
Fig.~\ref{fig:int_yield} results in an average value of $\alpha=1.38 \pm 
0.03 ({\rm stat}) \pm 0.07 ({\rm syst})$.

We have also considered the recently suggested scaling with the number of 
quark participants \Nqpart, which works well for charged particle 
production~\cite{ppg100}. Here \Nqpart is calculated with a Glauber Monte 
Carlo simulation similar to \Npart by picking random locations for 
constituent quarks within the nucleus. While our data is better described 
by scaling with a power-law in \Npart, it is also consistent with a 
power-law function $N_{qp}^\beta$, where $\Nqpart$ is the number of quark 
participants. In this case we find an exponent of $\beta=1.27 \pm 0.03 
({\rm stat}) \pm 0.07 ({\rm syst})$.

In most theoretical models thermal photon emission involves binary 
collisions of constituents, partons or hadrons, in hot and dense matter. 
Thus the emission rate from a unit volume should be proportional to the 
square of the number of constituents, while bulk particle production 
should scale with the number of constituents~\cite{Cerny, Kajantie}. Because 
particle production is approximately proportional to $N_{\rm part}$ one 
might expect thermal photon emission to scale as $N_{\rm part}^2$ 
times a correction for the increasing reaction volume with centrality. The 
increasing volume will reduce the centrality dependence, so that one 
expects $1 < \alpha < 2$ for thermal photon emission, just as observed.

Recent theoretical studies of the centrality dependence confirm our 
finding that the yield of thermal photon emission increases approximately 
with a power law function of \Npart. In the PHSD transport approach the 
power $\alpha$ is approximately 1.5~\cite{Linnyk:2013wma}, with no evident 
change in the shape of the spectra with centrality, very similar to our 
data. A hydrodynamic model~\cite{Shen:2013vja} shows a power law increase 
of the yield with a power $\alpha$ in the range from 1.67 to 1.9, 
increasing monotonically as the lower integration threshold increases from 
0.4 to 1.4\ GeV/$c$. Photon production in a glasma phase~\cite{Chiu:2012ij} 
was predicted to scale with $N_{\rm part}^\alpha$ with $1.47 < \alpha < 
2.2$. Other new production mechanisms, proposed to address the large 
$v_2$, have distinctly different centrality dependence. The yield from 
enhanced thermal photon emission in the strong magnetic field is expected 
to decrease with centrality, as the strength of the field weakens with 
decreasing impact parameter~\cite{Skokov}. The thermal photon yield should 
thus increase more slowly than expected from standard processes, but a 
quantitative estimate is not yet available.

\section{Summary and Conclusions}

We have isolated the low momentum direct photon yield emitted in \AuAu 
collisions.  The shape of the \pt spectra does not depend strongly on 
centrality, with an average inverse slope of ~240 MeV/$c$ in the range 
from 0.6 to 2 GeV/$c$.  The yield increases with centrality as 
$N_{\rm part}^\alpha$ with $\alpha\sim1.4$.  In conclusion, these 
results will help distinguish between different photon-production 
mechanisms and will constrain models of the space-time evolution of 
heavy ion collisions.


\section*{ACKNOWLEDGMENTS}


We thank the staff of the Collider-Accelerator and Physics
Departments at Brookhaven National Laboratory and the staff of
the other PHENIX participating institutions for their vital
contributions.  We acknowledge support from the 
Office of Nuclear Physics in the
Office of Science of the Department of Energy, the
National Science Foundation, Abilene Christian University
Research Council, Research Foundation of SUNY, and Dean of the
College of Arts and Sciences, Vanderbilt University (U.S.A),
Ministry of Education, Culture, Sports, Science, and Technology
and the Japan Society for the Promotion of Science (Japan),
Conselho Nacional de Desenvolvimento Cient\'{\i}fico e
Tecnol{\'o}gico and Funda\c c{\~a}o de Amparo {\`a} Pesquisa do
Estado de S{\~a}o Paulo (Brazil),
Natural Science Foundation of China (P.~R.~China),
Croatian Science Foundation and 
Ministry of Science, Education, and Sports (Croatia),
Ministry of Education, Youth and Sports (Czech Republic),
Centre National de la Recherche Scientifique, Commissariat
{\`a} l'{\'E}nergie Atomique, and Institut National de Physique
Nucl{\'e}aire et de Physique des Particules (France),
Bundesministerium f\"ur Bildung und Forschung, Deutscher
Akademischer Austausch Dienst, and Alexander von Humboldt Stiftung (Germany),
Hungarian National Science Fund, OTKA, K\'aroly R\'obert University College (Hungary), 
Department of Atomic Energy and Department of Science and Technology (India), 
Israel Science Foundation (Israel), 
National Research Foundation of Korea of the Ministry of Science,
ICT, and Future Planning (Korea),
Physics Department, Lahore University of Management Sciences (Pakistan),
Ministry of Education and Science, Russian Academy of Sciences,
Federal Agency of Atomic Energy (Russia),
VR and Wallenberg Foundation (Sweden), 
the U.S. Civilian Research and Development Foundation for the
Independent States of the Former Soviet Union, 
the Hungarian American Enterprise Scholarship Fund,
and the US-Israel Binational Science Foundation.



%
 
\end{document}